%% file: PilotMixtures.tex
\documentclass[10pt,twocolumn,journal]{IEEEtran}
\IEEEoverridecommandlockouts
\newcommand\hmmax{0}
\newcommand\bmmax{0}

\usepackage{lipsum}
\usepackage{graphicx}
\usepackage{tikz}
\usepackage{pgfplots}
\usepackage{xcolor}
\usepackage{color, colortbl}
\usepackage{amsmath}
\usepackage{bbm}
\usepackage{amsfonts, amssymb}
\usepackage{mathtools}
\usepackage{cite}
\usepackage[nolist]{acronym}

\usepackage{booktabs} 		% cute tables
\usepackage{diagbox}		% diagonal cell division in table
\usepackage{upgreek}
\usepackage{bbm}
\usepackage{Files/bm}
\usepackage{Files/winsnotation}
\usepackage{Files/Symbols_OPL}
\usepackage{Files/LVcolours} % cute colours

\usepackage[ruled,vlined,linesnumbered]{algorithm2e}
\usepackage{setspace}	 	% stretch on algorithm2e
\usepackage{comment}
\usepackage{array}
\newcolumntype{C}[1]{>{\centering\arraybackslash}p{#1}}

\makeatletter
\def\endthebibliography{%
  \def\@noitemerr{\@latex@warning{Empty `thebibliography' environment}}%
  \endlist
}
\makeatother
\IEEEoverridecommandlockouts
\usepackage{amsthm}
\theoremstyle{plain}

\theoremstyle{definition}

\pgfplotsset{compat=1.17}
\begin{document}
\pagestyle{empty}

\title{Exploiting Pilot Mixtures in Coded Random Access}

\author{%
  \IEEEauthorblockN{Lorenzo Valentini,~\IEEEmembership{Graduate~Student~Member,~IEEE,} Elena Bernardi,~\IEEEmembership{Graduate~Student~Member,~IEEE,} %Marco~Chiani,~\IEEEmembership{Fellow,~IEEE,} 
  Enrico~Paolini,~\IEEEmembership{Senior~Member,~IEEE}}
\thanks{The authors are with CNIT/WiLab, DEI, University of Bologna, Italy. Email: \{lorenzo.valentini13, elena.bernardi14, e.paolini\}@unibo.it. Supported by the European Union under the Italian National Recovery and Resilience Plan  of NextGenerationEU, partnership on ``Telecommunications of the Future'' (PE00000001 - ``RESTART'').}
%\thanks{The authors wish to thank Marco Chiani for his....}
}

% Don't write page number 0 to the cover page.
\maketitle 
%\markboth{Submitted to IEEE Transactions on Networking}{Rabbachin, Conti, and Win: Wireless Network Intrinsic Secrecy}

\input{Files/Acronimi_SICMMA.tex}
\setcounter{page}{1}

\thispagestyle{empty}

\begin{abstract}
The construction of preamble sequences for channel estimation by superposition of orthogonal pilots can improve performance of massive grant-free uplink from machine-type devices.
In this letter, a technique is proposed to obtain full benefit from these ``pilot mixtures'' in presence of a base station with a massive number of antennas. 
The proposed technique consists of combining pilot mixtures with an intra-slot successive interference cancellation (SIC) algorithm, referred to as inner SIC, to increase the number of decoded messages per slot.
In framed systems, the synergic effect of inner SIC and of an outer SIC algorithm across slots, typical of coded random access protocols, allows achieving a very high reliability with a low number of packet replicas per active user. 
%Massive MIMO - inner SIC - nested SIC
%\LVnote{\lipsum[1]} %occupies 14 rows
\end{abstract}

\begin{keywords}
Coded random access, grant-free access, massive MIMO, massive machine-type communication, 
successive interference cancellation.
\end{keywords}
%%%%%%%%%%%%%%%%%%%%%%%%%%%%%%%%%%%%%%%%%%%%%%%%%%%%%%
%%%%%%%%%%%%%%%%%%%%%%%%%%%%%%%%%%%%%%%%%%%%%%%%%%%%%%
\section{Introduction}
%%%%%%%%%%%%%%%%%%%%%%%%%%%%%%%%%%%%%%%%%%%%%%%%%%%%%%
%%%%%%%%%%%%%%%%%%%%%%%%%%%%%%%%%%%%%%%%%%%%%%%%%%%%%%
%\LVnote{1) Introduzione sul come in letteratura molti hanno cercato di usare tecniche di diversità per portare $G$ oltre 1 [Citare qualcosa che usa frenquenze diverse (forse LoRae?)]}

%[Origonale Elena-Lorenzo] Grant-free multiple access protocols for a massive number of devices have recently gained an increasing interest due to the new challenging key performance indicators envisaged for next-generation wireless networks \cite{Gui2020:6G,Kalalas2020:massive,Gao2021:MAC}. %Moreover, grant-free access eliminates the problem of the enormous amount of control signals that contending users generate to access the channel through a grant-based procedure.
Grant-free multiple access protocols have recently gained an increasing interest for the uplink of next-generation \ac{mMTC} applications, where an extremely large number of machine-type devices contend to deliver short packets to a common \ac{BS} \cite{Gui2020:6G,Kalalas2020:massive}. 
By eliminating handshake resource allocation procedures, that are typical of grant-based access, grant-free uplink protocols drastically reduce the amount of control signalling, making channel access very efficient in presence of  a massive number of devices with low duty cycle and unpredictable activity.
The resulting uplink protocol is very light on the device side, at the cost of an increased complexity in the receiver.
%for a massive number of devices  due to the new challenging key performance indicators envisaged for next-generation wireless networks . 
%[Originale Elena-Lorenzo] Aiming to next generation systems, we have that in \ac{mMTC} mild requirements as non trivial reliability and latency have to be taken into account within the system specifics \cite{Pokhrel2020:Towards}. Indeed, scalability is still the main metrics to compare schemes targeting \ac{mMTC} applications.
%It is currently agreed that n
Next-generation \ac{mMTC} use cases will impose more severe reliability and latency requirements, with respect to the ones typical of 5G, with scalability (i.e., the ability to support %connection of 
a very large number of devices) remaining the main key performance indicator \cite{Gao2021:MAC,Pokhrel2020:Towards}. %[Originale Elena-Lorenzo] To this aim, \ac{CRA} represent a good candidate to deal with this new challenge. In particular, the framed structure of typical \ac{CRA} schemes is able to imposed a fixed maximum latency, while the \ac{SIC} across the frame is able to achieve the required reliability.
In this context, \ac{CRA} schemes, combining resource diversity with \ac{SIC},
%and characterized by an inherent strong connection with codes-on-graphs, 
are emerging as candidates to achieve different scalability, reliability, and latency tradeoff points in a flexible manner (e.g., \cite{liva2011:irsa,paolini2015:magazine,Munari2021:age,Sor2018:coded,Valentini2022:SCACK}). 

%[Originale Elena-Lorenzo] To increase the number of decoded packets per slot, thus increasing scalability, some form of \ac{MPR} needs to be designed. A na{\"i}ve approach consists of investing in frequency resources. However, bandwidth is a precious resource and for this reason several works have chosen to exploit massive \ac{MIMO} technologies to achieve \ac{MPR}. One possible subdivision among them can be based on the preambles used for channel estimation, orthogonal \cite{Sor2018:coded, EmilLarsson19:NOMAOrto, Valentini2022:SCACK} or non-orthogonal \cite{Liu2018:sparse,Abebe2021:MIMO}. Recently, a preamble construction using orthogonal sequences was presented in \cite{Dai2021:PDRA}, for a single slot scenario. The constructed preambles are non-orthogonal to each other, but channel estimation processing exploits the orthogonality of the generating sequences. In particular, it was shown that using more pilots can help the reliability within a slot when grant-free scenario are considered. This idea can be interpreted as a diversity ALOHA \cite{Cho83:diversityALOHA} obtained through pilot diversity instead of slot diversity.

To increase the number of decoded packets per slot (i.e., scalability)
%(thus increasing scalability) 
at given reliability and latency, some form of \ac{MPR} needs to be exploited. 
While a naive approach consists of investing in frequency resources, more spectrally efficient ones are indeed possible. 
An example of them consists of letting machine-type devices share the same time-frequency resources, exploiting massive \ac{MIMO} processing at the \ac{BS} to achieve \ac{MPR}. 
Techniques in this area may be roughly classified based on the type of preambles used for channel estimation, namely, orthogonal \cite{Sor2018:coded, EmilLarsson19:NOMAOrto, Valentini2022:SCACK} versus non-orthogonal \cite{Liu2018:sparse,Abebe2021:MIMO}.
%[Originale Elena-Lorenzo] Recently, a new preamble construction exploiting orthogonal sequences was proposed in \cite{Dai2021:PDRA}, for a single slot scenario. The resulting preambles are non-orthogonal with respect to each other, but channel estimation processing exploits the orthogonality of the original sequences. In particular, it was shown that using more pilots can help the reliability within a slot when grant-free scenario are considered. This idea can be interpreted as a diversity ALOHA \cite{Cho83:diversityALOHA} obtained through pilot diversity instead of slot diversity.
Moreover, a new preamble construction was recently proposed that exploits the superposition of orthogonal pilots \cite{Dai2021:PDRA}.  
Although the resulting preambles are non-orthogonal with respect to each other, they still allow exploiting orthogonality of the building sequences. 
It turns out that mixing pilots helps reliability of grant-free access at single slot level. 
We point out that pilot superposition may also be interpreted as a form of diversity ALOHA \cite{Cho83:diversityALOHA} obtained through pilots instead of slots.

%[Originale Elena-Lorenzo] The principle of \ac{CRA} based on repetition codes \cite{liva2011:irsa, Munari2021:age, Ngo23:BAC}, is to gain in reliability and scalability by letting devices repeat their packets within a slotted frame, in a way similar to diversity ALOHA, and then perform \ac{SIC} across resources (e.g., time slots) to recover interfered users. This approach can be generalized to packet fragmentation and erasure coding, leading to \acl{CSA} \cite{paolini2015:csa}. Sticking to repetition-based \ac{CRA}, it was shown in \cite{Valentini2022:Joint} that, under realistic assumptions (fading channel and actual physical layer processing), the optimal \ac{CRA} should employ a constant repetition degree $2$ in an asymptotic setting. From an energy-saving perspective, this result shows that the optimal scheme in term of performance is also the optimal scheme in term of energy spent. This is due to the fact that the system is interference-limited. However, the error floor exhibited by scheme with repetition degree $2$ could not achieve the target reliability.

Repetition based \ac{CRA} schemes gain in reliability and scalability by letting devices repeat their packets within a frame, in a way similar to diversity ALOHA, and then perform \ac{SIC} across resources (e.g., time slots) to recover interfered users. 
This approach can be generalized to packet fragmentation and erasure coding, leading to \acl{CSA} \cite{paolini2015:csa}.
Sticking to repetition-based \ac{CRA}, it was shown in \cite{Valentini2022:Joint} that, as opposed to simpler (yet very popular) channel models such as the binary adder channel \cite{Ngo23:BAC}, under more realistic channel models and physical layer processing, the best \ac{CRA} configuration
%, as the frame size gets large and the device population size increases proportionally, 
is the one with a constant packet repetition degree $2$. 
From an energy-saving perspective, this result reveals that the optimal scheme in terms of performance is also optimal in terms of energy consumption. 
Unfortunately, this optimum scheme shows a high error floor in the \ac{PLR} curve which hinders its use at high target reliability values.

%\LVnote{3) In nuovi scenari \ac{mMTC} dove la reliability è importante bisogna introdurre tecniche di cancellazione per raggiungere le performance target. IN particolare noi cercheremo di capire in quali condizioni l'utilizzo di una Mixture of Pilots mi riesce a migliorare le performance in comparazione agli schemi in letteratura.}

%[Originale Elena-Lorenzo] Motivated by this result, in this letter we propose a novel \ac{CRA} scheme which effectively exploit more orthogonal sequences (a pilot mixture) to construct preambles. This scheme aims to make the optimal scheme derived in \cite{Valentini2022:Joint} appealing both in error floor and waterfall region.
Motivated by this result, in this letter we propose a new \ac{CRA} scheme that effectively exploits pilot mixtures and massive \ac{MIMO} processing to improve reliability, by means of a nested \ac{SIC} technique working both across pilots (at slot level) and across slots (at frame level).
Remarkably, the proposed technique makes the use of packet repetition degree $2$ appealing not only at high \ac{PLR} values but also at low ones, allowing full exploitation of the most energy efficient \ac{CRA} configuration.
%In fact, the possibility to use different slots for transmissions, alongside \ac{SIC} algorithms permits to achieve good reliability together with scalability also in grant-free contexts. 
The key contribution of this letter can be summarized as follows:
\begin{itemize}
    %[Originale Elena-Lorenzo] \item we propose a new nested \ac{SIC} procedure able to exploit pilot mixtures in a \ac{CRA} setting;
    \item A new nested \ac{SIC} procedure, able of exploiting pilot mixtures in the framework of \ac{CRA} schemes, is proposed;
    %[Originale Elena-Lorenzo] \item we discuss the \ac{CRA} scenarios in which pilot mixtures can provide benefits;
    \item The \ac{CRA} configurations in which pilot mixtures can provide benefits are discussed;
    %[Originale Elena-Lorenzo] \item we derive analytical performance bound.
    \item An analytical performance bound is derived to benchmark the actual performance.
\end{itemize}
%\LVnote{Forse l'organizzazione del paper si può togliere in una letter?}
% The paper is organized as follows. Section~\ref{sec:preliminary} introduces the system model and some preliminary concepts. 
% Section~\ref{sec:CRAwithPM} contains the proposed idea and analysis. %for \ac{MPR} scenarios. 
% Numerical results are provided in Section~\ref{sec:NumericalResults}. Finally, conclusions are drawn in Section~\ref{sec:Conclusions}.

\emph{Notation}: Capital and lowercase bold letters denote matrices and vectors, respectively. The conjugate transposition of a matrix or vector is denoted by $(\cdot)^H$, while $\|\cdot\|$ indicates the Euclidean norm. Furthermore, we denote the probability that a random variable $\rv{X}$ takes the value $x$, $\Pr(\rv{X} = x)$, as $P(x)$. 
Similarly, we write $P(x, y|z)$ in lieu of $\Pr(\rv{X} = x, \rv{Y} = y \,|\, \rv{Z} = z)$, and $P(\mathcal{E})$ to denote the probability that an event $\mathcal{E}$ occurs.

%%%%%%%%%%%%%%%%%%%%%%%%%%%%%%%%%%%%%%%%%%%%%%%%%%%%%%

%%%%%%%%%%%%%%%%%%%%%%%%%%%%%%%%%%%%%%%%%%%%%%%%%%%%%%
%%%%%%%%%%%%%%%%%%%%%%%%%%%%%%%%%%%%%%%%%%%%%%%%%%%%%%
\section{Preliminaries and Background}
\label{sec:preliminary}
%%%%%%%%%%%%%%%%%%%%%%%%%%%%%%%%%%%%%%%%%%%%%%%%%%%%%%
%%%%%%%%%%%%%%%%%%%%%%%%%%%%%%%%%%%%%%%%%%%%%%%%%%%%%%
In synchronous \ac{CRA}, the time is organized in frames, each 
%composed of 
with $N_\mathrm{s}$ slots. 
Each user, say user $k$, is slot- and frame-synchronous and contends for transmission of one information message. Time synchronization can be achieved exploiting a beacon broadcast by the \ac{BS} at the beginning of each frame.
%The user 
It forms a data payload $\V{x}(k) \in \mathbb{C}^{1\times N_{\mathrm{D}}}$
%of $N_{\mathrm{D}}$ complex symbols 
out of its message and 
%then 
sends multiple copies of 
the payload on the frame.
Each of these transmitted packets, 
%or ``bursts'', 
features its own preamble and a copy of the payload and fits exactly one slot. 
%To build its multiple bursts,
%To this aim, 
%Each active user independently draws a repetition degree $r$ at random, according to a probability distribution $\Lambda_r$. Then, it randomly draws $r$ slot indexes and, for each such slot, it draws a pilot repetition degree $p$ according to a probability distribution $\Lambda_p$.
Each active user draws randomly, independently of the other users: 
(i) a repetition degree $r$ with probability distribution $\Lambda_r$; 
(ii) $r$ slot indexes, uniformly and without replacement; 
(iii) a preamble order $p$, with probability distribution $\Psi_p$, independently in each chosen slot.
Let $\{\V{s}_1, \dots, \V{s}_{N_\mathrm{P}}\}$ be a set of $N_\mathrm{P}$ orthogonal pilots. 
Then, in each chosen slot the user builds a ``pilot-mixture'' preamble sequence combining $p$ pilots in that set
%$\{\V{s}_1, \dots, \V{s}_{N_\mathrm{P}}\}$ 
uniformly without replacement, similar to \cite{Dai2021:PDRA}. 
In the generic such slot, the preamble built by user $k$ is 
%given by 
%
\begin{align}\label{eq:PilotMixture}
    \V{p}(k) = \frac{1}{\sqrt{p}} \sum_{j = 1}^{p}\V{s}(k,j)\,,
\end{align}
where $\V{s}(k,j) \in \{\V{s}_1, \dots, \V{s}_{N_\mathrm{P}} \}$ is the $j$-th 
%sequence 
pilot drawn by user $k$, and the transmitted packet is $[\V{p}(k),\V{x}(k)]$.
Hereafter, we adopt the \acl{PGF} notation $\Lambda(x) = \sum_r \Lambda_r x^r$ and, similarly, $\Psi(x)= \sum_p \Psi_p x^p$ \cite{liva2011:irsa}. 
%When the distribution is concentrated
For concentrated distributions, we simply use the parameter $r$ or $p$, e.g., $r = 2$ stands for $\Lambda(x) = x^2$.

Packets are sent over a Rayleigh block fading channel with coherence time equal to the slot time. 
Perfect power control is assumed.
%At \ac{PHY} layer, 
At the receiver side, the \ac{BS} features a massive number of antennas, $M$. 
The signal received by the \ac{BS} in a slot may be expressed as $[\M{P}, \M{Y}] \in \mathbb{C}^{M \times (N_\mathrm{P}+N_\mathrm{D})}$ where
\begin{align}\label{eq:PAndY}
    \M{P} = \sum_{k \in \mathcal{A}} \V{h}_k \V{p}(k) + \M{Z}_p \quad \text{and} \quad \M{Y} = \sum_{k \in \mathcal{A}} \V{h}_k \V{x}(k) + \M{Z} .
\end{align}
In \eqref{eq:PAndY}: $\mathcal{A}$ is the set of users
%transmitting 
with a packet in the slot under analysis; $\V{h}_k = (h_{k,1}, \dots, h_{k,M})^T \in \mathbb{C}^{M\times 1}$ is the channel coefficient vector for user $k$ in the %considered 
slot, whose elements are \ac{i.i.d.} random variables with distribution $\mathcal{CN}(0, \sigma_\mathrm{h}^2)$ for all $k \in \mathcal{A}$; $\M{Z}_p \in \mathbb{C}^{M\times N_\mathrm{P}}$ and $\M{Z} \in \mathbb{C}^{M\times N_\mathrm{D}}$ are matrices of Gaussian noise samples.
%; $\V{p}(k) \in \mathbb{C}^{1\times N_\mathrm{P}}$ and $\V{x}(k) \in \mathbb{C}^{1\times N_\mathrm{D}}$ are the preamble and payload transmitted by user $k$ in the considered slot. 
Due to power control, $\sigma_\mathrm{h}^2$ is the same for all users and without loss of generality we can assume $\sigma_\mathrm{h}^2 = 1$.

% The processing is split into two phases. In phase~$1$, all slots are processed in order. 
%%%%%%%%%%%%%%%%%%%%%%%%%%%%%%%%%%%%%
% In each slot, the \ac{BS} first attempts channel estimation for all possible pilots by computing $\V{\phi}_j\in \mathbb{C}^{M\times 1}$ for all $j \in \{1,\dots,N_{\mathrm{P}}\}$, as
% %
% \begin{align}
% \label{eq:PhiEstimate}
%     \V{\phi}_j &= \frac{\M{P} \,\V{s}_j^{H}}{\| \V{s}_j \|^2} = \sum_{k \in \mathcal{A}^j} \V{h}_k + \V{z}_j\,.
% \end{align}
% %
% Here, $\mathcal{A}^j$ is the set of active users employing pilot $j$ in the current slot, $\V{s}_j \in \mathbb{C}^{1\times N_\mathrm{P}}$ is the $j$-th pilot sequence, and $\V{z}_j \in \mathbb{C}^{M \times 1}$ is a noise vector.
% Then, the \ac{BS} attempts payload estimation using \ac{MRC} as
% \begin{align}
% \label{eq:PayloadEst}
%     \hat{\V{x}}_j = \frac{\V{\phi}_j^{H} \, \M{Y}}{\| \V{\phi}_j \|^2}\,.
% \end{align}
%%%%%%%%%%%%%%%%%%%%%%%%%%%%%%%%%%%%%%%%%%%%%%%%%
%[Originale Elena-Lorenzo] Within a slot, the \ac{BS} attempts channel estimation and payload estimation for all pilots, by computing the vectors $\V{\phi}_j\in \mathbb{C}^{M\times 1}$ and $\hat{\V{x}}_j \in \mathbb{C}^{1\times N_{\mathrm{D}}}$, for all $j \in \{1,\dots,N_{\mathrm{P}}\}$, as
At the \ac{BS}, \ac{MPR} capability can be obtained exploiting massive \ac{MIMO} and the set of orthogonal pilots, $\{\V{s}_1, \dots, \V{s}_{N_\mathrm{P}}\}$, similar to \cite{Sor2018:coded}.
In each slot, the \ac{BS} attempts channel and payload estimation for all pilots $\V{s}_j$, $j \in \{1,\dots,N_{\mathrm{P}}\}$, by computing the vectors $\V{\phi}_j\in \mathbb{C}^{M\times 1}$ and $\hat{\V{x}}_j \in \mathbb{C}^{1\times N_{\mathrm{D}}}$ as
\begin{align}
\label{eq:PhiAndPayEstimate}
    \V{\phi}_j &= \frac{\M{P} \,\V{s}_j^{H}}{\| \V{s}_j \|^2} = \sum_{k \in \mathcal{A}^j} \V{h}_k + \V{z}_j\,\,\,\, \mathrm{and} \,\,\,\,  \hat{\V{x}}_j = \frac{\V{\phi}_j^{H} \, \M{Y}}{\| \V{\phi}_j \|^2}\,,
\end{align}
where $\mathcal{A}^j \subseteq \mathcal{A}$ is the subset of users employing pilot $j$ in the current slot 
%$\V{s}_j \in \mathbb{C}^{1\times N_\mathrm{P}}$ is the $j$-th pilot sequence, 
and $\V{z}_j \in \mathbb{C}^{M \times 1}$ is a noise vector.
Due to pilot orthogonality, 
%neglecting noise, the channel estimate $\V{\phi}_j$ coincides with $\V{h}_k$ whenever, in the considered slot, $\V{s}_j$ appears only in the pilot-mixture preamble of user $k$. 
$\V{\phi}_j$ is a maximum likelihood estimate of $\V{h}_k$ whenever, in the considered slot, $\V{s}_j$ appears only in the pilot-mixture preamble of user $k$.
%On the other hand, due to superposition of the data payloads of different users we have $\hat{\V{x}}_j \neq \V{x}_k$ also when in absence of noise.
%However, due to statistical quasi-orthogonality of the channel vectors, also known as favorable propagation effect, we have $\hat{\V{x}}_j \simeq \V{x}_k$ for large $M$ \cite{Val23:SICAlgo}.
Moreover, due to statistical quasi-orthogonality of the channel vectors (also known as favorable propagation) we have $\hat{\V{x}}_j \simeq \V{x}(k)$ for large $M$, which makes decoding of $\V{x}(k)$ possible \cite{Val23:SICAlgo}.

%%%%%%%%%%%%%%%%%%%%%%%%%%%%%%%%%%%%%%%%%%%%%%%%%%%%%%

\begin{figure}[t]
    \centering
    % \resizebox{0.925\columnwidth}{!}{
    %     \input{Figures/scenario.tex}
    % }
    \includegraphics[width=0.925\columnwidth]{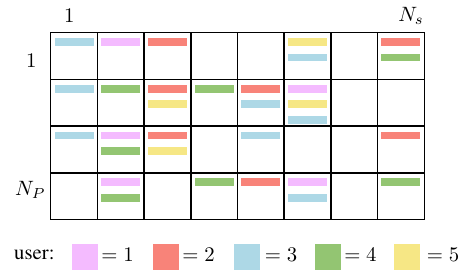}
    \caption{Pictorial representation of the chosen pilots in a frame. In the example, the frame is composed by $N_\mathrm{s} = 8$ slots; $N_\mathrm{P} = 4$ orthogonal pilots are available to $5$ active devices, with $\Lambda(x) = 0.5 x^2 + 0.5 x^3$ and $\Psi(x) = 0.5 x^2 + 0.5 x^3$, meaning that each user has $50\%$ probability to pick $2$ or $3$ slots, and for each slot has a $50\%$ probability to choose $2$ or $3$ pilots.}
    \label{fig:frameGrid}
\end{figure}

%%%%%%%%%%%%%%%%%%%%%%%%%%%%%%%%%%%%%%%%%%%%%%%%%%%%%%
%%%%%%%%%%%%%%%%%%%%%%%%%%%%%%%%%%%%%%%%%%%%%%%%%%%%%%
\section{Coded Random Access with Pilot Mixtures} 
\label{sec:CRAwithPM}
%%%%%%%%%%%%%%%%%%%%%%%%%%%%%%%%%%%%%%%%%%%%%%%%%%%%%%
%%%%%%%%%%%%%%%%%%%%%%%%%%%%%%%%%%%%%%%%%%%%%%%%%%%%%%

The access scheme described in Section~\ref{sec:preliminary} can be pictorially represented by a grid, as in Fig.~\ref{fig:frameGrid}. 
Orthogonality among packets in different slots (or columns) is guaranteed by time division, while packets placed in the same slot but using different pilots (rows) are separable, as per the above processing, only when the total number of active users in the slot, $K_\mathrm{s}$, is small compared to $M$. 
It should be noted that slot diversity implies an energy cost, while pilot diversity comes at no extra transmit energy due to the normalization in \eqref{eq:PilotMixture}. 
For clarity, in Fig.~\ref{fig:frameGrid}, placing packets in different column cells has not the meaning of repeating packets, but it means that those pilots were picked by the same user. 
The interpretation as a repetition, however, is a useful analogy from a \ac{CRA} perspective.

To get full benefit from the resources offered by the frame grid, we propose a \emph{nested} \ac{SIC} mechanism. 
%In particular, 
We aim at obtaining a performance gain by combining diversity and \ac{SIC}, not only at a frame level across different slots, but also at a slot level by means of pilot diversity achieved by \eqref{eq:PilotMixture} %and multipath diversity achieved 
exploiting massive \ac{MIMO}. % and orthogonal pilots. 
Hereafter, we refer to this latter intra-slot \ac{SIC} as ``inner \ac{SIC}''.
With reference again to Fig.~\ref{fig:frameGrid}, this means performing \ac{SIC} both across columns, as conventionally done in \ac{CRA}, and across rows, enabled by pilot mixtures.
In every slot, the \ac{BS} processes the received samples \eqref{eq:PAndY}, by performing \eqref{eq:PhiAndPayEstimate} for all $j \in \{1,\dots,N_{\mathrm{P}}\}$.
Successful channel decoding on $\hat{\V{x}}_j$, for any $j$, triggers the inner \ac{SIC} procedure in the current slot.
The whole packet is therefore reconstructed using the information provided by the payload\footnote{Using the payload bits as a seed of a random number generator (shared between users and \ac{BS}), the receiver can trace back all choices the user has made for preamble construction and packet frame placements.}, and its interference is subtracted from the signal samples as
\begin{equation}
\begin{aligned}\label{eq:PYmod}
\M{P}^{(i+1)} = \M{P}^{(i)} - \V{\phi}_j \V{p} \quad\text{and}\quad 
\M{Y}^{(i+1)} = \M{Y}^{(i)} - \V{\phi}_j \V{x}\,,
\end{aligned}
\end{equation}
where $\V{p}$ and $\V{x}$ are the reconstructed preamble and payload of the successfully decoded user.
The index $i$ indicates the number of \ac{SIC} operations done in the current slot, with initial conditions $\M{P}^{(0)} = \M{P}$ and $\M{Y}^{(0)} = \M{Y}$.
Note that \eqref{eq:PYmod} relies on the channel estimate $\V{\phi}_j$, since $\V{\phi}_j$ is not interfered by other users due to orthogonality. 
It is thus not necessary to re-estimate the channel coefficients in this phase.
After any interference subtraction operation, the cleaned signal samples are processed again by \eqref{eq:PhiAndPayEstimate} and possibly \eqref{eq:PYmod}, until no more packets can be decoded.
Information of any decoded packet is stored in a buffer for outer \ac{SIC} processing.
The proposed slot-by-slot processing is summarized in Algorithm~\ref{algo:InnerSIC}. To reduce complexity, activity detection procedures can be adopted to process only slot-pilot pairs where users may be retrieved.

\begin{algorithm}[t]
%\DontPrintSemicolon
\caption{Slot-by-slot processing with inner SIC}\label{algo:InnerSIC}
\ForAll{$n \in \{1, \dots, N_\mathrm{s}\}$}{
\ForAll{$j \in \{1, \dots, N_\mathrm{P}\}$}{
    \textbf{Channel Estimation}: compute $\V{\phi}_j$ as in \eqref{eq:PhiAndPayEstimate}\;
    \textbf{Payload Estimation}: compute $\hat{\V{x}}_j$ as in \eqref{eq:PhiAndPayEstimate}\;
    \If{$\hat{\V{x}}_j$ contains a valid packet}{
        Reconstruct the packet preamble $\V{p}$\;
        \textbf{Inner SIC}: update $\M{P}$ and $\M{Y}$ as in \eqref{eq:PYmod}\;
        Store the packet for \textbf{outer SIC} processing\;
        Set $j=1$ to restart the slot computation\;
    }
}
}
\end{algorithm}

At the end of the frame, once the \ac{BS} has concluded the inner \ac{SIC} phase, the receiver enters the outer \ac{SIC} phase. This is the typical \ac{CRA} \ac{SIC} across slots \cite{paolini2015:csa}. 
Here, buffered decoded packets are processed in order. For each such packet, the contribution of replica interference is subtracted from the corresponding slots and decoding is reattempted in these slots for all pilots.
%where the receiver processes the buffer of decoded packets element-by-element to subtract the interference from the frame and attempts to recover new users. 
Since in block fading channel the channel gains vary in each slot, it is necessary to estimate the channel for each replica to be subtracted. 
%rising the problem of channel estimation during the replica interference cancellations. 
Solutions to this issue have been proposed, e.g., in \cite{Sor2018:coded} and \cite{Val23:SICAlgo}. We consider the payload aided based algorithm \cite{Val23:SICAlgo}, estimating the channel as
\begingroup
\allowdisplaybreaks
\begin{align}
\label{eq:hTilde}
    \widehat{\V{h}}^{(i)} &= \frac{\M{Y}^{(i)} \,\V{x}^{H}}{\| \V{x} \|^2}
\end{align}
\endgroup
and subtracting the interference on the stored slot symbols as
\begin{equation}
\begin{aligned}\label{eq:PYmodTilde}
    \M{P}^{(i+1)} = \M{P}^{(i)} - \widehat{\V{h}}^{(i)} \V{p} \quad \text{and} \quad
    \M{Y}^{(i+1)} = \M{Y}^{(i)} - \widehat{\V{h}}^{(i)} \V{x}\,,
\end{aligned}
\end{equation}
%
%where, again, $\V{p}$ and $\V{x}$ are the reconstructed preamble and payload of the previously decoded user which had been stored in the buffer. 
where $\V{p}$ and $\V{x}$ are the preamble and payload of the buffered packet. 
%After the subtraction of the interference, for each pilot $j$, the decoding is re-attempted according to \eqref{eq:PhiAndPayEstimate} on the current $\M{P}^{(i)}$ and $\M{Y}^{(i)}$.
%Then, whenever a new user is successfully decoded its information is stored at the end of the buffer and the outer \ac{SIC} phase is iterated until the buffer contain no more users. 
Whenever a new packet is successfully decoded, its information is pushed in the buffer. 
The outer \ac{SIC} phase is iterated until the buffer becomes empty.

\subsection{Benefits of Exploiting Pilot Mixtures}\label{subsec:PM}
%The effective exploitation of pilot mixtures in a framed system is not trivial. To this aim, we analyze how this technique can be employed in a \ac{CRA} protocol to improve the performance compared to the state-of-the-art and which benefits it achieves. 

To describe the benefits of the proposed technique, let us firstly analyze an intrinsic source of error in \ac{CRA} schemes with power control. 
In this regime, an unresolvable collision occurs when two or more users pick the same pilot mixtures in the same slots.
Focusing only on this source of error allows us computing a lower bound on the \ac{PLR}, $P_\mathrm{L}$, useful for parameter design when inner \ac{SIC} is employed. 
Restricting our analysis on concentrated distributions $\Lambda(x) = x^r$ and $\Psi(x) = x^p$, we have that the probability that at least two users are involved in such a collision event is
\begin{align}\label{eq:Pu}
    P_{\mathrm{u}} &= 1 - \prod_{i=0}^{N - 1} \frac{C- i}{C}\,,
\end{align}
where $C$ is the total number of possible choices that users can make and $N$ is the number of active users.
The exact meaning of $N$ depends on the context as explained next.
%is the active population size. 
In a slotted but unframed system we have $C = \binom{N_\mathrm{P}}{p}$ and $N = K_\mathrm{s}$, where $K_\mathrm{s}$ is the number of active devices per slot. Moreover, in a framed and slotted system with nested \ac{SIC} we have $C = \binom{N_\mathrm{s}}{r} \, \binom{N_\mathrm{P}}{p}^r$ and $N = K_\mathrm{a}$, where $K_\mathrm{a}$ is the number of active devices per frame.
Finally, we can lower bound the \ac{PLR} as $P_\mathrm{L} \ge 2\, P_{\mathrm{u}} \, / \, N$, since at least two users out of $N$ are involved in the unresolvable collision.
%where $2/N$ is the best case \ac{PLR} when a collision occurs.
%accounting for a double error whenever this event occurs.
This lower bound results tight for small $K_\mathrm{s}$ (or $K_\mathrm{a}$), i.e., in the low traffic regime.
%due to the fact that the considered error is predominant in the low traffic region.

The combination of pilot mixtures with \ac{SIC} can bring several benefits to the access protocol. 
The first advantage this technique bears, consists of a reduced error floor. In fact, in \cite{Valentini2022:Joint} it was shown that, differently to what was proven under collision channel, the best performance is obtained with $\Lambda(x) = x^2$, or equivalently for $r=2$.
However, for $p=1$ this distribution exhibits high error floors which limit its adoption. 
In contrast, for $p>1$ (pilot mixtures), consistently with \eqref{eq:Pu}, the error floor appears at remarkably lower \ac{PLR} values.
A second advantage is that, adopting the sole inner \ac{SIC}, the scheme is still able to achieve a remarkable performance as it will be highlighted in Section~\ref{sec:NumericalResults}. 
This can be appealing if we need to process data only in real-time and without storing the whole frame at the \ac{BS}. 
This also reduces the processing burden the \ac{BS} should manage. 
A third advantage can be achieved in presence of \ac{ACK} messages at the end of each slot as proposed in \cite{Valentini2022:SCACK}. 
The usage of \ac{ACK} messages permits to save user energy due to transmission interruption. 
In this case, the inner \ac{SIC} used in presence of pilot mixtures is able to retrieve more users per slot compared to conventional schemes using $p=1$. 
Moreover, more \acp{ACK} imply less interference in the frame, which also improves performance.

\subsection{Benchmarks without Successive Interference Cancellation}

Let us start with a slotted but unframed scheme where users exploit pilot mixtures but no inner \ac{SIC} is performed at the receiver. 
The performance depends essentially on two factors: $i)$ the probability of ``singleton'' packets, in which at least one pilot of the mixture is chosen by only one user; % which are the only ones having a particular pilot in the slot); 
$ii)$ The successful decoding probability of singleton packets. 
We can see that $i)$ depends on the access protocol, while $ii)$ is strictly related to the \ac{PHY} layer and channel model. 
For example, in Fig.~\ref{fig:frameGrid}, no user $5$ packets are singleton ones, while user $1$ has a singleton in slot $2$.
Assuming that a singleton always triggers a successful packet decoding, given $K_\mathrm{s}$ users per slot, the \ac{PLR} is tightly approximated as
\begin{align}
\label{eq:PLS}
    P_\mathrm{L}^{(\mathrm{S})} \approx \sum_p \left( 
    1 - \left( 
    1 - \frac{\Psi^\prime(1)}{N_\mathrm{P}}
    \right)^{K_\mathrm{s}-1}
    \right)^p \Psi_p\,.
\end{align}
%
% \begin{align}
% \label{eq:PLS}
%     P_\mathrm{L}^{(\mathrm{S})} = \left( 
%     1 - \left( 
%     1 - \frac{p}{N_\mathrm{P}}
%     \right)^{K_\mathrm{s}-1}
%     \right)^p\,.
% \end{align}
To derive \eqref{eq:PLS}, let us focus on an active user with preamble order $p$ in the slot. 
Decoding of the user's packet fails whenever all pilots in its mixture are picked by other users. 
The probability that any pilot in the mixture is chosen by another device with a preamble order $p$ is $\binom{N_\mathrm{P}-1}{p-1} / \binom{N_\mathrm{P}}{p} = p/N_\mathrm{P}$.
%the event in which one of the selected pilot is collided by an interfered, $\mathcal{I}$, given its pilot repetition rate $p$ is $p/N_\mathrm{P}$. 
This probability, averaged over the preamble order of the interferer, becomes
$\Psi^\prime(1) / N_\mathrm{P}$.
%By the law of total probability, we have that $P(\mathcal{I}) = \Psi^\prime(1) / N_\mathrm{P}$. 
Then, the probability that one of the chosen pilots is also picked by at least one interferer is $1 - \left( 1 - \Psi^\prime(1) / N_\mathrm{P} \right)^{K_\mathrm{s}-1}$. 
Treating the collision events for all $p$ pilots in the mixture as independent, 
%independence of the collision events for all $p$ pilots in the mixture, 
and averaging with respect to the preamble order distribution leads to the approximation in \eqref{eq:PLS}.
%We consider independence between the failure events of each chosen pilot. 
%Finally, considering that the user under examination has picked $p$ sequences and exploiting again the law of total probability we obtain \eqref{eq:PLS}. 
Note that \eqref{eq:PLS} holds with equality for any concentrated distribution $\Psi(x) = x^p$; for irregular distributions it matches very tightly the actual simulated results, with negligible deviations only for small $K_\mathrm{s}$\footnote{Equation \eqref{eq:PLS} assumes successful packet decoding when the packet is a singleton one. In this respect, it represents a lower bound on the actual \ac{PLR}.
It is possible to account also for $ii)$ in the analysis, i.e., \ac{PHY} layer impairments and channel model, as done in \cite{Val23:SICAlgo}. For a large number of antennas, the impact of $ii)$ is however negligible compared to $i)$.}.
In Section~\ref{sec:NumericalResults} we will show the beneficial impact of the inner \ac{SIC} on the overall performance, which extends the range of load values for which a small \ac{PLR} can be attained.

The above reasoning can be extended to a slotted and framed scheme where users exploit slot diversity and pilot mixtures, but no form of \ac{SIC} (neither inner nor outer) is performed at the receiver. 
Given that there are $K_\mathrm{a}$ devices active on the frame and assuming a user message is not successfully received when all pilots in its mixture are interfered in all $r$ chosen slots,
%Without \ac{SIC} mechanisms, 
we obtain the \ac{PLR} approximation
\begin{align}
\label{eq:PLneverSIC2}
    P_\mathrm{L}^{(\mathrm{F})} \approx &\sum_r \Lambda_r \Bigg[
    \sum_{p} \Psi_p \sum_{j=0}^{p} (-1)^j \binom{p}{j} \notag \\
    &\left(
    1 - \frac{\Lambda^\prime(1)}{N_\mathrm{s}} +
    \frac{\Lambda^\prime(1)}{N_\mathrm{s}} \left( 
    1 - \frac{\Psi^\prime(1)}{N_\mathrm{P}}
    \right)^j
    \right)^{K_\mathrm{a}-1}
    \Bigg]^r\,.
\end{align}
%\begin{proof}
In fact, reasoning as before, the \ac{PLR} can be approximated as $\sum_r (1 - P(\mathcal{U}))^r \Lambda_r$, where $P(\mathcal{U})$ is the probability that a replica has at least one singleton pilot. %uncollided. 
Using \eqref{eq:PLS}, we have that the probability of the event $\mathcal{U}$ given that $K_\mathrm{I} = K_\mathrm{s} - 1$ interfering users are present in the slot is $P(\mathcal{U}|K_\mathrm{I}) = 1 - P_\mathrm{L}^{(\mathrm{S})}$, while $K_\mathrm{I}$ is binomial distributed with success probability $\Lambda^\prime(1)/N_\mathrm{s}$ and $K_\mathrm{a} - 1$ trials.
Finally, by the law of total probability we have that $P(\mathcal{U}) = \sum_{K_\mathrm{I} = 0}^{K_\mathrm{a}-1} P(\mathcal{U}|K_\mathrm{s}) \, P(K_\mathrm{s})$ which yields \eqref{eq:PLneverSIC2}.
Similar to \eqref{eq:PLS}, independence assumptions among pilot and replica collision events make \eqref{eq:PLneverSIC2} an approximation on the actual \ac{PLR} for irregular distributions.
On the other hand, for concentrated distributions $\Lambda(x) = x^r$ and $\Psi(x) = x^p$, \eqref{eq:PLneverSIC2} becomes exact and holds with equality.

%%%%%%%%%%%%%%%%%%%%%%%%%%%%%%%%%%%%%%%%%%%%%%%%%%%%%%
%%%%%%%%%%%%%%%%%%%%%%%%%%%%%%%%%%%%%%%%%%%%%%%%%%%%%%
\section{Numerical Results}
\label{sec:NumericalResults}
%%%%%%%%%%%%%%%%%%%%%%%%%%%%%%%%%%%%%%%%%%%%%%%%%%%%%%
%%%%%%%%%%%%%%%%%%%%%%%%%%%%%%%%%%%%%%%%%%%%%%%%%%%%%%
%%%%%%%%%%%%%%%%%%%%%%%%%%%%%%%%%%%%%%%%%%%%%%%%%%%%%%
\subsection{Simulation Setup}
\label{subsec:SimSetUp}

Simulation results are provided for \ac{CRA} schemes %proposed in Section~\ref{sec:Contributions}, 
in which each user transmits data payloads encoded with an $(n = 511, k = 421, t = 10)$ binary \ac{BCH} code. 
Part of the $k$ information bits are used to validate the decoded packets via a \ac{CRC}.
After padding the \ac{BCH} codeword with a final zero bit, the encoded bits are mapped onto a \ac{QPSK} constellation with Gray mapping, yielding a data payload of $N_\mathrm{D}=256$ symbols. 
Simulation results are given for 
%symbol rate $B_\mathrm{s}=1\,\mathrm{Msps}$ and 
$M = 256$ \ac{BS} antennas.
Pilots are constructed using Hadamard matrices \cite{Stuber2001:Book}.
Both schemes with and without feedback are considered for the numerical analysis.
In case a feedback channel is used, it is assumed ideal 
%The numerical results assume a perfect feedback channel for \acp{ACK} 
(i.e., all \acp{ACK} are always successfully received). 
We plot the \ac{PLR} $P_\mathrm{L}$ against the number of active users, representing the system scalability parameter.
This latter parameter is equal to $K_\mathrm{s}$ for a slotted and unframed system, while it is equal to $K_\mathrm{a}$ for a slotted and framed one.
In the framed case, we use $N_\mathrm{s} = 62$ slots.

\begin{figure}[t]
    \centering
    % \resizebox{0.99\columnwidth}{!}{
    %     \input{Figures/graph_singleSlot_v2.tex}
    % }
    \includegraphics[width=0.9\columnwidth]{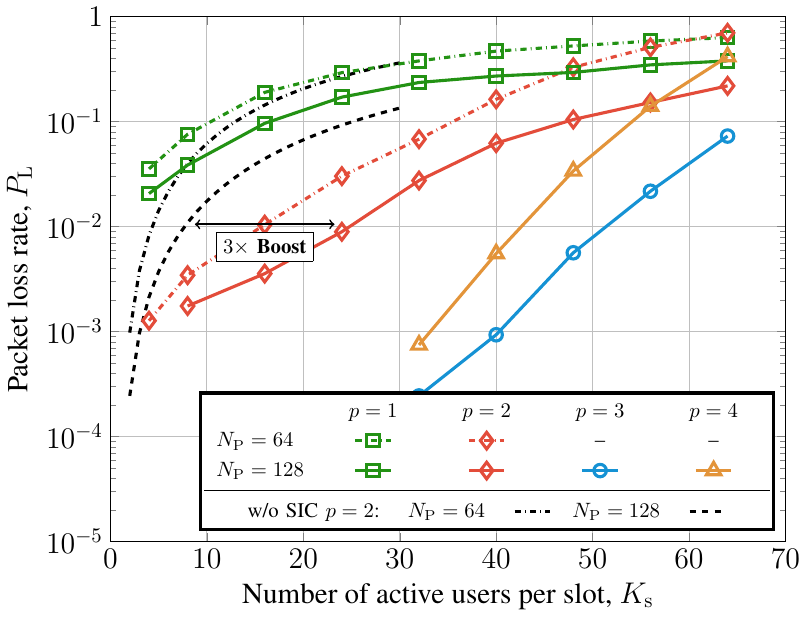}
    \caption{Packet loss rate comparison between different pilot mixture schemes in a slotted and unframed scenario.}
    \label{fig:slot}
\end{figure}

%%%%%%%%%%%%%%%%%%%%%%%%%%%%%%%%%%%%%%%%%%%%%%%%%%%%%%
\subsection{Performance Evaluation}
\label{subsec:NumRes}
\begin{figure}[t]
    \centering
    % \resizebox{0.99\columnwidth}{!}{
    % \input{Figures/graph_PAB_pilot_mixture_frame_v2.tex}
    % }
    \includegraphics[width=0.9\columnwidth]{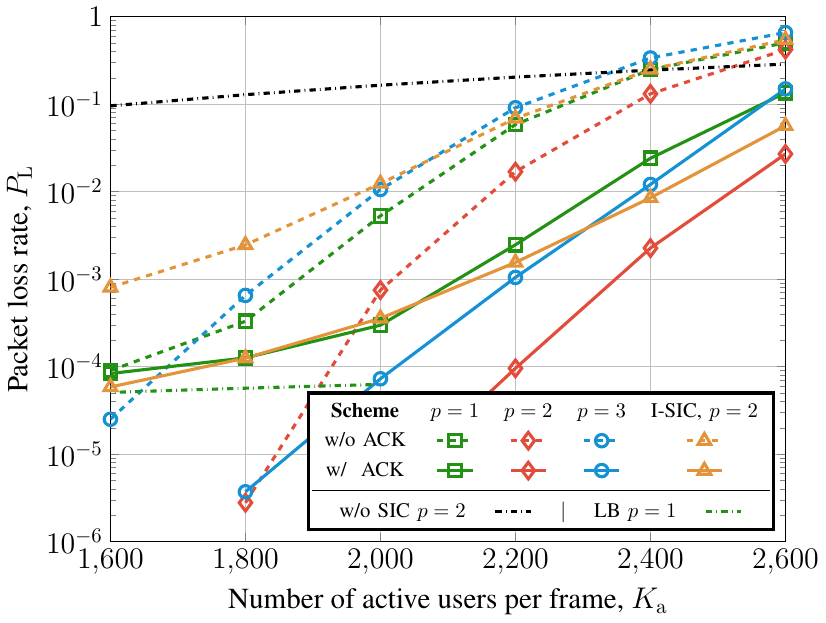}
    \caption{Packet loss rate comparison between different pilot mixture schemes in a framed scenario. I-\ac{SIC}: inner \ac{SIC}.}
    \label{fig:frame}
\end{figure}

\emph{1) Slotted and Unframed System Analysis:} 
%\subsubsection{Slotted and Unframed System Analysis}
%Fig.~\ref{fig:slot} shows the \ac{PLR} 
%the probability that the information of an active user is not received correctly by the BS (i.e., the \ac{PLR})
%versus the total number of active users in a single slot, $K_\mathrm{s}$. 
%Specifically, 
In Fig.~\ref{fig:slot}, we compare schemes using pilot mixtures and inner \ac{SIC} ($p > 1$) varying the number of available orthogonal pilot sequences $N_\mathrm{P} \in \{ 64, 128\}$, for a slotted but unframed system. 
%The packets placed in the same slot can use different number of pilots to construct the preamble, $p = 1, 2, 3$.
The figure also shows the comparison between different concentrated distributions $\Psi(x) \in \{ x, x^2, x^3, x^4\}$.
%characterized by different repetition pilots rates $p = 1, 2, 3, 4$.
The gain attained over the scheme with $\Psi(x) = x$ can be interpreted as the performance boost given by pilot mixtures and proposed inner \ac{SIC}.
%is a schemwe report a \ac{CRA} scheme without pilot mixtures (i.e., using $p = 1$).
In the figure, we emphasize that increasing the preamble order $p$ has a positive effect for small values of $p$ (e.g., moving from $p=1$ to $p=3$ in the example).
In contrast, when $p$ increases beyond a certain value, pilot overload in the slot deteriorates the \ac{PLR} due to impossibility of performing accurate channel estimation, an effect that becomes predominant.
%In fact, when $p$ increases, the \emph{virtual} traffic, represented by the number pilots chosen in that slot, increases. 
%As a consequence, this causes more probable pilot collisions as depicted in figure comparing the curve with $p=3$ and $p=4$.
A second result we emphasize is the effectiveness of the inner \ac{SIC} in comparison to the analytical and ideal curve without \ac{SIC} derived in \eqref{eq:PLS}. 
For example, considering $N_\mathrm{P} = 128$ and $p=2$, inner \ac{SIC} increases by a factor of $3$ %is able to triple 
the number of served users per slot at a target \ac{PLR} $P_\mathrm{L}^* = 10^{-2}$.
We also note that increasing $p$ in absence of inner \ac{SIC} does not significantly improve scalability at low \ac{PLR}.
Finally, we can see that increases $N_\mathrm{P}$, the performance improves as expected.\\
%
%\subsubsection{Slotted and Framed System Analysis} 
\indent\emph{2) Slotted and Framed System Analysis:} 
In Fig.~\ref{fig:frame} we analyze the role of the preamble order $p$ in a slotted and framed system.
We report the \ac{PLR} for schemes using: $i)$ pilot mixtures and the proposed nested \ac{SIC}; $ii)$ pilot mixtures and only inner \ac{SIC}; $iii)$ no pilot mixture (i.e., $p=1$) and outer \ac{SIC}.
Results are carried out also allowing the possibility to employ \ac{ACK} messages at the end of each slot, as suggested in \cite{Valentini2022:SCACK}. 
The reception of an \ac{ACK} by an active user enables the possibility to preemptively stop transmissions of subsequent replicas by the user, that saves energy and does not generate unnecessary interference in future slots.
A comparison is provided varying the number of active users in the frame, $K_\mathrm{a}$, for different concentrated distributions $\Psi(x) \in \{ x, x^2, x^3\}$. 
For all schemes we adopt $r=2$ (i.e., $\Lambda(x) = x^2$) and $N_\mathrm{p} = 128$.
Using \eqref{eq:PLneverSIC2} we observe that we cannot achieve good scalability at low \ac{PLR} without \ac{SIC}. 
This is due to the fact that we are exploiting only time diversity, while \ac{SIC} permits to retrieve several collided users.
As an example, for $p=2$ we obtain a target $P_\mathrm{L}^* = 10^{-3}$ at $K_\mathrm{a} \approx 400$ without \ac{SIC}.
Enabling only inner \ac{SIC}, we significantly improve scalability to $K_\mathrm{a} \approx 1600$ for the same $P_\mathrm{L}^*$.
Next, enabling \ac{ACK} messages we obtain a further scalability gain, reaching $K_\mathrm{a} \approx 2150$. 
Lastly, with the proposed nested \ac{SIC} we end up with  $K_\mathrm{a} \approx 2350$.
As reviewed in Section~\ref{subsec:PM}, the state-of-the-art schemes using $p=1$ may suffer from intolerably high error floors under the asymptotically optimal distribution $\Lambda(x) = x^2$ \cite{Valentini2022:Joint}. 
Pilot mixtures represent a valid solution to lower the error floor without sacrificing the waterfall performance. 
In this example, fixing $K_\mathrm{a} = 1800$ and using the analysis in Section~\ref{subsec:PM}, the \ac{PLR} of the scheme with $p=1$ is lower bounded by $P_\mathrm{L} \ge 5.7 \cdot 10^{-5}$, while that of the scheme with $p=2$ is lower bounded by $P_\mathrm{L} \ge 1.4 \cdot 10^{-8}$.
%In fact, error floors can be a problem also when below the target \ac{PLR} due to ``soft junction'' between the waterfall region and the error floor region. 
%Although the green and blue curve are similar in waterfall region, the gap between them tends to increase roughly one decade above the error floor. 
Unexpectedly, as opposed to the unframed case,  %(Fig.~\ref{fig:slot})
we highlight that in the framed one the choice $p=2$ outperforms $p=3$. 
%This is in contrast to what show in Fig.~\ref{fig:slot} and 
This reveals that focusing on single slot analysis may provide inaccurate results and conclusions in framed system, which motivates \ac{CRA} design and analysis.
%%%%%%%%%%%%%%%%%%%%%%%%%%%%%%%%%%%%%%%%%%%%%%%%%%%%%%
%%%%%%%%%%%%%%%%%%%%%%%%%%%%%%%%%%%%%%%%%%%%%%%%%%%%%%
\section{Conclusions}
\label{sec:Conclusions}
A grant-free \ac{CRA}-type access scheme has been proposed that exploits both massive \ac{MIMO} and pilot mixture preambles through a nested \ac{SIC} algorithm. 
As a remarkable result, the proposed processing is able to substantially lower the high error floor typical of energy-efficient \ac{CRA} configurations featuring two replicas per user. 
Analytical performance benchmarks have been derived to assess the improvement due to \ac{SIC}, as well as a lower bound useful for error floor estimation and system design.
The proposed approach provides good performance even in slotted but unframed systems, where only the inner \ac{SIC} algorithm is applied, which is useful in situations where low-complexity \ac{BS} processing is required.

%%%%%%%%%%%%%%%%%%%%%%%%%%%%%%%%%%%%%%%%%%%%%%%%%%%%%%
%%%%%%%%%%%%%%%%%%%%%%%%%%%%%%%%%%%%%%%%%%%%%%%%%%%%%%

%%%%%%%%%%%%%%%%%%%%%%%%%%%%%%%%%%%%%%%%%%%%%%%%%%%%%%
%%%%%%%%%%%%%%%%%%%%%%%%%%%%%%%%%%%%%%%%%%%%%%%%%%%%%%
%%%%%%%%%%%%%%%%%%%%%%%%%%%%%%%%%%%%%%%%%%%%%%%%%%%%%%
%%%%%%%%%%%%%%%%%%%%%%%%%%%%%%%%%%%%%%%%%%%%%%%%%%%%%%

\section*{Acknowledgements}
We would like to thank M. Chiani for helpful
discussions.
% This work has been carried out in the framework of the CNIT National Laboratory WiLab and the WiLab-Huawei Joint Innovation Center.

%%%%%%%%%%%%%%%%%%%%%%%%%%%%%%%%%%%%%%%%%%%%%%%%%%%%%%
%%%%%%%%%%%%%%%%%%%%%%%%%%%%%%%%%%%%%%%%%%%%%%%%%%%%%%
%\bibliographystyle{IEEEtran}
%\bibliography{Files/IEEEabrv,Files/StringDefinitions,Files/StringDefinitions2,Files/refs}

%%%%%%%%%%%%%%%%%%%%%%%%%%%%%%%%%%%%%%%%%%%%%%%%%%%%%%
%%%%%%%%%%%%%%%%%%%%%%%%%%%%%%%%%%%%%%%%%%%%%%%%%%%%%%

\end{document}

%% file: Files/Acronimi_SICMMA.tex
\begin{acronym}
% usage: \ac{SW}, \acp{SW} for plurals \acf{SW} Use the full name of the acronym.
%\acs{SW}Use the acronym, even before the first corresponding \ac command
%\acl{acronym}Expand the acronym without using the acronym itself.
\small
\acro{ACK}{acknowledgement}
\acro{AWGN}{additive white Gaussian noise}
\acro{BCH}{Bose–Chaudhuri–Hocquenghem}
\acro{BN}{burst node}
\acro{BS}{base station}
\acro{CDF}{cumulative distribution function}
\acro{CRA}{coded random access}
\acro{CRC}{cyclic redundancy check}
\acro{CRDSA}{contention resolution diversity slotted ALOHA}
\acro{CSA}{coded slotted ALOHA}
\acro{eMBB}{enhanced mobile broad-band}
\acro{FER}{frame error rate}
\acro{IFSC}{intra-frame spatial coupling}
\acro{i.i.d.}{independent and identically distributed}
\acro{IRSA}{irregular repetition slotted ALOHA}
\acro{LB}{lower bound}
\acro{LDPC}{low-density parity-check}
\acro{LOS}{line of sight}
\acro{MAC}{medium access control}
\acro{MIMO}{multiple input multiple output}
\acro{ML}{maximum likelihood}
\acro{MMA}{massive multiple access}
\acro{mMTC}{massive machine-type communication}
\acro{MPR}{multi-packet reception}
\acro{MRC}{maximal ratio combining}
\acro{PAB}{payload aided based}
\acro{PDF}{probability density function}
\acro{PGF}{probability generating function}
\acro{PHY}{physical}
\acro{PLR}{packet loss rate}
\acro{PMF}{probability mass function}
\acro{PRCE}{perfect replica channel estimation}
\acro{QPSK}{quadrature phase-shift keying}
\acro{RF}{radio-frequency}
\acro{SC}{spatial coupling}
\acro{SIC}{successive interference cancellation}
\acro{SIS}{successive interference subtraction}
\acro{SN}{sum node}
\acro{SNB}{squared norm based}
\acro{SNR}{signal-to-noise ratio}
\acro{URLLC}{ultra-reliable and low-latency communication}
\end{acronym}